\documentclass{article}

\PassOptionsToPackage{numbers, compress}{natbib}

\usepackage[final]{neurips_2022}
\usepackage[utf8]{inputenc} 
\usepackage[T1]{fontenc}    
\usepackage{hyperref}       
\usepackage{url}            
\usepackage{booktabs}       
\usepackage{amsfonts}       
\usepackage{nicefrac}       
\usepackage{microtype}      

\usepackage{caption}
\usepackage{subcaption}
\usepackage{amsthm}
\newtheorem{definition}{Definition}
\usepackage{multicol}

\title{Indexing AI Risks with Incidents, Issues, and Variants}

\author{%
  Sean McGregor\\
  Responsible AI Collab.\\
  California\\
  \texttt{hai@seanbmcgregor.com}
  \And
  Kevin Paeth\\
  MIT CSAIL\\
  Massachusetts\\
  \texttt{paeth@mit.edu}
  \And
  Khoa Lam\\
  Responsible AI Collab.\\
  Vietnam\\
  \texttt{lkchemposer@gmail.com}
}

\begin{document}

\maketitle

\begin{abstract}
Two years after publicly launching the AI Incident Database (AIID) as a collection of harms or near harms produced by AI in the world, a backlog of ``issues'' that do not meet its incident ingestion criteria have accumulated in its review queue. Despite not passing the database's current criteria for incidents, these issues advance human understanding of where AI presents the potential for harm. Similar to databases in aviation and computer security, the AIID proposes to adopt a two-tiered system for indexing AI incidents (i.e., a harm or near harm event) and issues (i.e., a risk of a harm event). Further, as some machine learning-based systems will sometimes produce a large number of incidents, the notion of an incident ``variant'' is introduced. These proposed changes mark the transition of the AIID to a new version in response to lessons learned from editing 2,000+ incident reports and additional reports that fall under the new category of ``issue.''
\end{abstract}

In November of 2020, the AIID publicly launched as a collection of harms and near harms realized in the real world involving intelligent systems \cite{Mcgregor2021}. Now two years later, the database has expanded to more than 300 incidents characterized by more than 2,000 incident reports. These indexing activities have engendered considerable discussion around incident reporting as a potential mechanism to bring about transparency, accountability, and responsible governance of AI systems \cite{AIA2021,nistrmp,brief2021ai}. As the AIID organizes various policy, corporate, and researcher stakeholder groups to collect and share AI incidents, concept definitions play an imperative role in setting standards and common understanding. As AI incident reporting is a fast developing practice, definitions must simultaneously be implementable in future mandatory and voluntary reporting frameworks \cite{AIA2021}, and accommodate uncertain and incomplete reporting about incidents. To meet these requirements, we propose a shift from a single tier (i.e., only index incidents), to a database containing ``incidents'' and ``issues'' -- where issues are a more permissively defined category. Our proposed \textit{short} definitions are stated as,



\begin{definition}[AI Incident]
An alleged harm or near harm event where an AI system is implicated.
\end{definition}
\begin{definition}[AI Issue]
An alleged harm by an AI system that has yet to occur or be detected.
\end{definition}

These two definitions draw inspiration from the Common Vulnerabilities and Exposures (CVE) system \cite{TheMITRECorporation2020}, whereby an event (e.g., exposure of a web server hosting private data) is treated separately from a risk (a known vulnerability of the web server library). Exposures serve to motivate other parties to respond to vulnerabilities, but interested parties should not require the example of an exposure to take appropriate mitigating steps. Similarly, AI issues serve as notices that an AI incident is likely to occur if the community of AI system deployers does not take mitigating steps.

Tiered systems also present flexibility for different reporting requirements. Aviation safety databases \cite{ntsb} separate a higher tier of ``accidents'' from a lower tier of ``incidents'' based on whether there was injury, death, or sufficient damages. In the United States, aviation incident reporting is voluntary unless compelled by an interested authority, whereas accidents must always be reported \cite{FAA_AIM}. As regulatory bodies move to require AI incident disclosure within their jurisdictions, a similar dichotomy is likely to develop with different jurisdictions shifting the boundary between incidents and issues. Further, in the event countries do enact penalties for failures to disclose AI incidents, the existence of a reporting tier that does not presume incident status provides incentives (e.g., waiver of penalties) for disclosure first, and sorting out the classification later.

Without a second tier, settling the boundary between ``incident'' and database rejection is controversial. The AIID received at least nine reports of facial recognition systems deployed by Israeli security forces at Palestinian checkpoints. These reports are yet to be ingested as incidents, pending a known misidentification of a Palestinian by facial recognition. At that time the accumulated reports can objectively be added to a new incident. Alternatively, the AIID could treat the system's existence as impacting the Palestinian people's rights to privacy and non-discrimination. When taking this framing, Palestinians may \textit{also} argue that the removal of the facial recognition system from checkpoints may still not remove the harms imposed by the human operators of the checkpoint -- this candidate incident may fail the requirement that the incident not occur in the absence of an AI system.
Categorizing these reports as AI issues would prevent these salient potential harms from being totally excluded from the database.

Assume another challenging checkpoint scenario: in the event that all misclassifications by the facial recognition model are reported, should all misidentification events be considered different incidents? Indexing each incident takes editor time and triggers downstream investigation and classification effort. Human capacities to index incident data will quickly be overwhelmed if manual processes are required for every discrete event. Herein lies a tension between indexing incidents at depth with added insights and indexing incidents at breadth to cover all occurrences of substantially similar incidents. We address this tension with the introduction of incident ``variants'' defined as,

\begin{definition}[AI Incident Variant]
An incident that shares the same causative factors, produces similar harms, and involves the same intelligent systems as a known AI incident.
\end{definition}

The variant concept accommodates the multiplicity of events that are likely to repeat, reducing the human labor required for subsequent event processing. In the checkpoint example, each misidentification of a Palestinan by these systems would result in an incident variant because it results from the same system, goals, technical failings, harms, and harmed population.

Variants also extend to model types that produce thousands of unreported incidents. Text-to-text (e.g., GPT-3 \cite{brown2020language}) and text-to-image (e.g., Stable Diffusion \cite{Stability}) models produce incident variations with the same causative factors and similar harms. These variants can be collected under the first incident sharing similar factors and inform subsequent investigation.

While solving the multiplicity problem, variants introduce an identification problem. Incidents currently in the AIID can be cited via their permanent URLs (e.g., \textit{incidentdatabase.ai/cite/1}), but the concept of ``variant'' makes it difficult to cite specific events. Here we chose to expand the incident identifier,

\begin{definition}[Incident Identifier]
A universally unique identifier for the incident and its variants ({NAMESPACE:INCIDENT\#.VARIANT\#}). The ``namespace'' refers to the numbering authority,\footnote{Currently only ``AIID'', but likely others (e.g., country codes) in the future} the ``incident'' refers to the AI incident number, and the ``variant'' refers to the incident variant.
\end{definition}


This identifier is backwards compatible since single number identifiers are assumed to be the originally reported incident. Further, it presumes a future where incident databases are federated and built to be interoperable via the introduction of the namespacing. Already a variety of jurisdictions are considering developing country-specific incident databases (e.g., follow on efforts from \cite{AIA2021}). Efforts must be made to ensure the hard won lessons within each country are broadly shared. Starting with a standard by which incidents are namespaced ensures incidents can continue to be referenced and indexed without ambiguity.

These proposed changes result both from the experience indexing incidents and operational necessities stemming from a changing policy environment. While most incident reports are presently derived from popular press journalism of AI misbehavior, we believe a world of high-volume incident reporting requires terminology and processes to capture misbehavior comprehensively, collaboratively, and practically. It is by advancing our sense of history that we may build and govern a human-centered future with AI.

\pagebreak
\textbf{Acknowledgements}

Kate Perkins gave valuable feedback on the contents of this work in addition to her duties as an AIID incident editor. The definitions and discussions presented within the paper are also greatly influenced by ongoing efforts by the Organisation for Economic Co-operation and Development (OECD) to adopt a shared definition of AI incident across all 38 member states. Finally, the AIID is an effort by many people and organizations organized under the banner of the Responsible AI Collaborative, including the Center for Security and Emerging Technology (CSET), whose Zachary Arnold contributed to the first incident criteria and definition. It is through the Collab's collective efforts that the ontological perspectives presented above have meaning and real world importance.

\bibliography{references}




\end{document}